\documentclass[twocolumn,showpacs,amsmath,amssymb,pr,superscriptaddress]{revtex4-1}
\usepackage{graphicx}
\usepackage{epsf}
\usepackage{ulem}
\usepackage{color}
\usepackage[unicode=true,colorlinks=true,urlcolor=blue,citecolor=blue]{hyperref}

\begin{document}

\title{Low Temperature Electrical Properties of CVD Graphene on LiNbO$_3$: Acoustic Studies
}

\author{I. L. Drichko}
\author{I. Yu. Smirnov}
\affiliation{Ioffe Institute, Russian Academy of Sciences, St. Petersburg, 194021 Russia}
\author{Yu. M. Galperin}
\affiliation{Department of Physics, University of Oslo, P. O. Box 1048 Blindern, Oslo, 0316 Norway}
\affiliation{Ioffe Institute, Russian Academy of Sciences, St. Petersburg, 194021 Russia}
\author{P.A. Dementev}
\affiliation{Ioffe Institute, Russian Academy of Sciences, St. Petersburg, 194021 Russia}
\author{M.G. Rybin}
\affiliation{Prokhorov Institute of General Physics of the Russian Academy of Sciences, 119991 Moscow, Russia}

\begin{abstract}
Contactless acoustic methods were used to determine electrical parameters - electrical conductivity, carrier mobility and their concentration - in single-layer graphene deposited on the surface of lithium niobate.
\end{abstract}


\maketitle

\section{Introduction}

Graphene is the unique two-dimensional material of XXI century. Its main advantage is unprecedented carriers mobility, caused by presence of free electron on p-orbital, that is delocalized along the whole  graphene  monolayer  and can move with high  velocity  as  massless  fermion.  Due to this particular unique property of graphene  the  bright future was predicted for its application in nanoelectronics.  However,  in  actual  practice  it   turned   out that it is very hard to reach the record-high values  of  carriers mobility in graphene,  let alone the manufacturing  of  samples,  that  can  be  reproduced  on  an  industrial scale. Graphene quality  and  possibility  of  its  application in nanoelectronics are usually characterized with charge carriers mobility in  this  material.  Thus,  the  task  of  quality control by means of analysis of charge carriers mobility  in  graphene  is  of  great  relevance.     The  highest
values of charge carriers mobility,  up  to  $10^6$~cm$^2$/Vs, were  achieved  in  freely  suspended  quasineutral  graphene,
produced by the method of epitaxial growth on silicon carbide  in  vacuum  at  low  temperatures~\cite{1}.  However,  this method of graphene producing is poorly scaled and
expensive. For  mass  commercial  synthesis  of  graphene the method of chemical vapor deposition (CVD) of  graphene on copper foil surface is used. This method
is easier to implement and more cost effective than the epitaxial growth. Charges  mobility  in  graphene  produced by CVD method is lower, but  also  can  be  high  under certain conditions, reaching the values of up to
 $\mu = 3.5 \times  10^5$~cm$^2$/Vs~\cite{2}.

For  material  optimizations  it  is  very  important  to have
the reliable procedure of mobility measurement. The most popular approach for charges mobility measurement is the production of a field transistor with graphene channel~\cite{3}.

This method allows to make measurements with high accuracy, but preparation to measurements requires more efforts, particularly lithography and etching of graphene, as well  as  metal  contacts  deposition  on  graphene.   Another
method (4-contact, as  per  Van  der  Pauw  in  magnetic  field~\cite{4}) is an alternative option of mobility measurement. However,  the  probes  contacting  with  sample  can  result in
sample damage.

The third method of mobility measurement is a terahertz spectroscopy~\cite{5}.  This is a non-contact  method,  allowing  to define the charges mobility in graphene. The method
allows to study the distribution of conductivity and mobility over the sample surface. But sensitivity of the method is relatively low ($\sigma <$ 0.1~mS).

In  this  study  we  use  non-contact   method   of   charges
mobility measurement in graphene at low  temperatures using analysis of distribution of surface acoustic waves (SAW) along the interface of piezodielectric (LiNbO$_3$)
and   graphene.      Simultaneous   measurement   of velocity
and attenuation of SAW allows to define the complex conductance of graphene on SAW frequency (see, for instance, review~\cite{6}).

In this study the results of measurement of two graphene samples, produced using the method of chemical vapor deposition on copper foil surface, and passed to piezodielectric LiNbO$_3$ are demonstrated. Using the above mentioned
technique~\cite{6} the electrophysical properties of  graphene were   studied  and  its  charge   carriers  mobility  was calculated. Graphene samples, different in surface morphology, that is caused by peculiarities of graphene synthesis and possibility to control the process with high accuracy, were used  in  the   study.    Samples   are   different   in   copper foil temperature during synthesis, particularly, differences
in temperature  by  10--20$^{\text o}$C  between  the  samples  result in  variation   of  grains   quantity  in  polycrystal   sample of
graphene and, as a result, various charges scattering on their boundaries. The procedure  of  measurements  is  described in the study in detail, and differences in electrophysical parameters of samples with various surface morphology are described.

\section{Samples and experimental procedure}

In this study we examined two graphene samples, produced using the method of  chemical  vapor  deposition on copper foil surface from gas mixture of argon, hydrogen and methane at temperature of 850$^{\text o}$C and reduced pressure
of 100 mbar.  The main differential characteristic of graphene
synthesis method  is  a  method  of  copper  foil  heating  with direct passage of current through it.  This  method allows to control the heating and cooling rate with high accuracy, while the temperature is measured by pyrometer
through inspection window in vacuum chamber. Detailed information in synthesis method is presented in studies~\cite{7,8}. Disadvantages of this method include small temperature gradient of copper foil of 10-20$^{\text o}$C at a distance of 10 mm (in sample with dimensions of 20$\times$20 mm$^2$ the temperature varies from 850$^{\text o}$C in  the  center  of  the  sample  to  830$^{\text o}$C at  sample  edges). During  study  we  used  one  synthesized
sample with dimensions of 20$\times$20 mm$^2$, divided into several parts with dimensions of 7$\times$10 mm$^2$, two of which were passed to piezoelectric (lithium niobate). In this study the sample 1 had the reduced temperature of 830$^{\text o}$C during synthesis,  while  sample  2  had  the  temperature  of  850$^{\text o}$C
during synthesis. Both graphene samples were passed from copper foil to lithium niobate surface using the standard
method of "wet" transition with polymethylmethacrylate as
a supporting polymer and ammonium persulphate as an
etchant for copper.

Piezoelectric properties and high quality of LiNbO$_3$ crystals give significant advantages for studying the two-dimensional and quasi-two-dimensional materials. Particularly, strong piezoelectric effect of LiNbO$_3$ allows to use oscillating electrical fields, generated by surface acoustic wave
(SAW) at its propagation over the piezocrystal and entering the low-dimensional system, placed on its surface~\cite{9}. As a result, the absorption, $\Gamma$, and velocity, $v$, of SAW depend on
the electrical characteristics of the surface layer and can be used for quantitative determination of the latter. This is the main idea behind the acoustic method of low-dimensional
materials analysis, that we used for various systems (see for review, e.g.,~\cite{7}).

Three essential advantages make this technique rather prospective.

(1)	The method does not require electrical contacts, therefore the results do no depend on them.

(2)	Electronic contributions to absorption and SAW velocity depend on magnetic field, and these dependencies
allow to separate the electronic contributions from lattice contributions.

(3) Relation of dynamic high-frequency (HF) conductivity, absorption and sound velocity includes values, measured experimentally.

\begin{figure}[h]
\centerline{
\includegraphics[width=9cm,clip=]{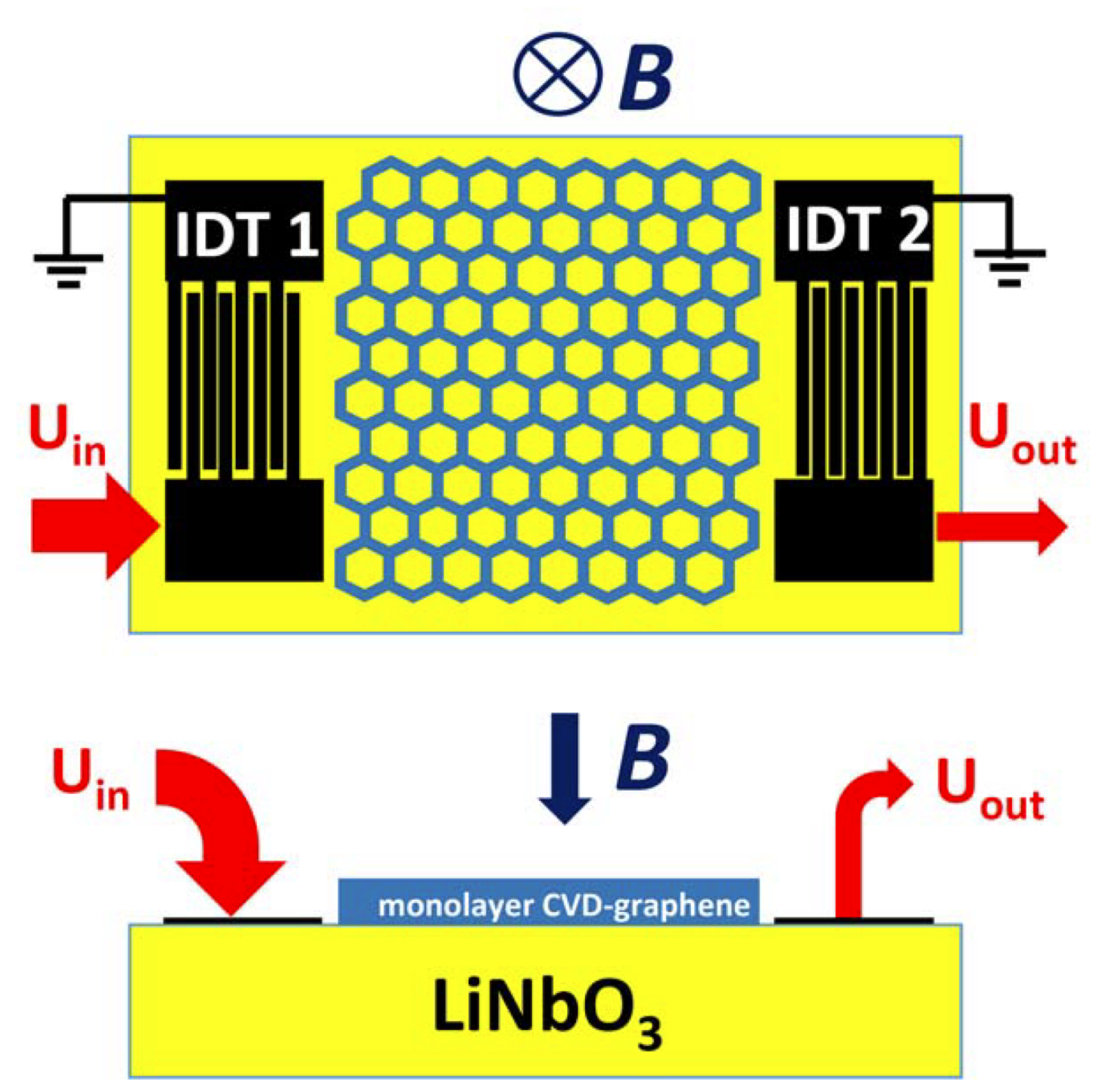}}
\caption{Scheme of acoustic experiment. $U_{in}$ and $U_{Uout}$ are the input and output signals, $B$ is the magnetic field.} \label{fig1}
\end{figure}

For this experiment the monolayer graphene film was deposited on polished surface of lithium niobate, on which the converters IDT 1 and IDT 2 (of gold) were pre-formed for generation and receiving the surface acoustic waves. Figure~\ref{fig1} demonstrates the experimental scheme.  SAW  frequencies were 28, 85,  140,  197  and 252 MHz (odd  harmonics of the main converter frequency of 28 MHz). Holder with sample was put to cryostat with superconducting magnet and cooled to temperatures of 1.7-4.2 K. Absorption and
variation of SAW  velocity of various frequencies depending
on magnetic field of up to 8 T were measured during the experiment.

Absorption and variation of SAW velocity  propagation are defined by a sum of two contributions caused by crystal lattice and charge carriers.  Since  the  lattice contributions  in non-magnetic materials do not depend on magnetic field, for separation of these contributions it is naturally to use the dependencies of acoustic characteristics on magnetic field.

In approximation, linear to SAW amplitude, the contributions of charge carriers are defined with complex dynamic conductance, $\sigma (\omega, \textbf{k}) \equiv \sigma_1(\omega, \textbf{k}) - i \sigma_2(\omega, \textbf{k})$, that
generally depends on frequency of SAW $\omega$, and its wave
vector $\textbf{k}$. Relation between SAW propagation characteristics and conductance is the following (see, for instance,~\cite{6}):

\begin{equation}
\label{GV}
\begin{gathered}
\Gamma= k  \frac{K^2}{2}  \frac{\sigma_1/\sigma_M}{[(1 + \sigma_2/\sigma_M)^2 + (\sigma_1/\sigma_M)^2 ]},\\
\Delta v/v = \frac{K^2}{2}\frac{1 + \sigma_2/\sigma_M}{[(1 + \sigma_2/\sigma_M)^2 + (\sigma_1/\sigma_M)^2]}.
\end{gathered}
\end{equation}
Here, $k \equiv |\textbf{k}|$, $K^2 /2 = 2.25\times 10^{-2}$ is the electromechanical constant of lithium niobate, $\sigma_{1,2}$ are the real  and  imaginary parts of dynamic conductivity $\sigma (\omega)$ per square (conductance); $\sigma_M = v_0 \times (\varepsilon_1 + \varepsilon_{gr})$,  $v_0$  is the SAW  velocity  at $B$=0; $\varepsilon_1$, $\varepsilon_{gr}$ are dielectric constants of lithium niobate and graphene, respectively. During calculations we used  values $\varepsilon_1$=50, $\varepsilon_{gr}$=6.9~\cite{10}.

From  the simultaneously measured  absorption and  variation of SAW velocity the complex conductance of graphene can be calculated,
$\sigma (\omega) = \sigma_1 (\omega) - i \sigma_2 (\omega)$. From analysis of these values it can be concluded that $\sigma_1$ and $\sigma_2$ weakly
depend on frequency over the whole examined frequency range, while $\sigma_2(\omega) \ll \sigma_1(\omega)$. Such relation supports the metal nature of graphene layer conductivity, while the dynamic conductance $\sigma_1(\omega)$ is close to static conductance
$\sigma (B) \equiv \sigma_{xx} (\omega = 0, B)$.

It  should  be  noted  that  in  the  examined  system  the
graphene layer is a polycrystalline material with significant electrons scattering over grain boundaries. This is indicated by, particularly, measured conductivity value. We suppose that such material is not characterized with Dirac cones anymore, but can be rather characterized as a system of conductivity electrons with concentration n and quadratic
spectrum at effective mass $m^*$ and momentum relaxation time $\tau$.

In non-quantizing magnetic field the conductivity of such surface layer is defined as per formula
\begin{equation}
\label{sB}
\sigma (B) = \sigma_{0}/[1+(\omega_c \tau)^2],
\end{equation}
where $\sigma_{0}$ is the conductivity in the absence of  magnetic field, $\omega_c$ is the cyclotron frequency. When using Drude equation  for  electrons  with  isotropic  quadratic  spectrum,
$\omega_c = eB/m^*c$, where $c$ is the light velocity. In strong magnetic field, when the following condition is met
\begin{equation}
\label{wc}
(\omega_c \tau)^2=(\mu B/c)^2 \gg 1,
\end{equation}
where $\mu$ is the charge carriers mobility, $\sigma (B) \propto B^{-2}$. If  we build the dependence of the experimentally defined conductivity on $1/B^2$, and it is linear, from this linear
dependence we can define the slope $A$ and relation
\begin{equation}
\label{muc}
(\mu /c)^2=\sigma_0/A.
\end{equation}

If we take a point on the experimental curve  $\sigma (B)$, where the strong field condition (\ref{wc}) is not met, after simple calculations we get
\begin{equation}
\label{s0}
\sigma_0=\sigma (B)/[1-\sigma (B) B^2/A].
\end{equation}

Thus, determination of electronic characteristics using acoustic method is as follows:

1)	building of the experimental dependence $\sigma (1/B^2)$ in the area of strong magnetic fields, calculation of the slope $A$ of linear dependence $\sigma = A/B^2$;

2)	determination of $\sigma (0)$ as per formula (\ref{s0});

3)	determination of mobility as per formula (\ref{muc}) and current concentration as per formula
\begin{equation}
\label{n}
n=\sigma (0)/e \mu.
\end{equation}

This method of graphene characteristics determination is rather approximate, since we know only the approximate graphene film length ($\approx$7 mm). Besides, limited reproducibility of the results of various experiments on the same
sample also impacts the accuracy.

\section{Experimental results}

In the study we examined 2 samples, with monolayer CVD graphene, deposited on lithium niobate surface using the same method, described above. However, the samples characteristics are significantly different. Measurements of absorption $\Gamma$ and $\Delta v/v$ were  made in temperature interval
of (1.7-4.2) K, in frequency range of (28-300) MHz in magnetic fields of up to 8 T. Temperature variation of $\Gamma$ and
$\Delta v/v$ in zero magnetic field was studied for sample 2 at frequencies of 30 and 140 MHz.

\subsection{SAW absorption at $B$ = 0}

In sample 1 the large absorption of SAW, $\simeq$30~dB/cm, that grows with temperature decrease, is observed even at room temperature.

\begin{figure}[h]
\centerline{
\includegraphics[width=8cm,clip=]{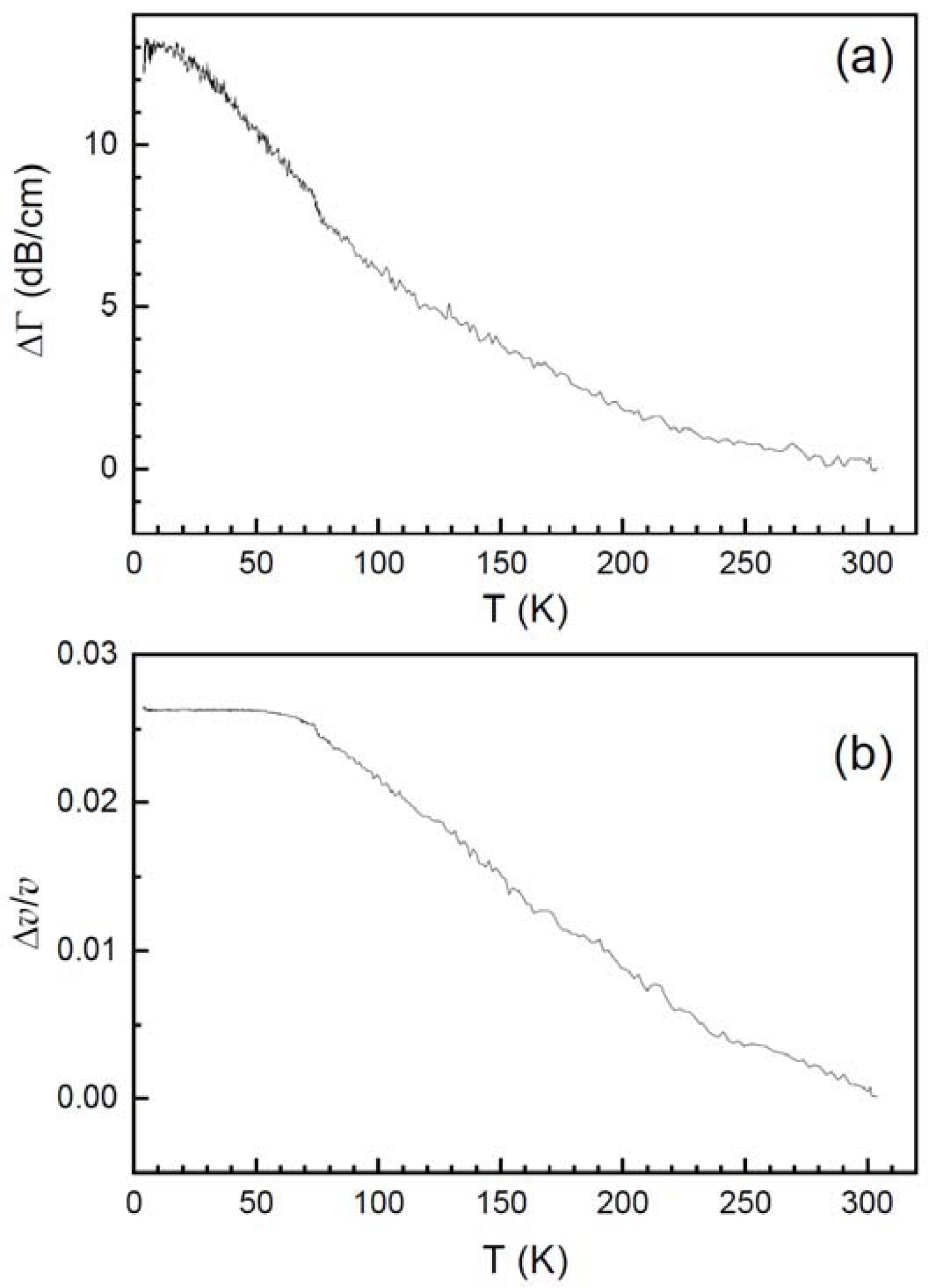}}
\caption{Dependence (a) of absorption coefficient $\Delta\Gamma$ and (b)
$\Delta v/v$ on temperature $T$ ; $f$=30~MHz, $B$=0~T.} \label{fig2}
\end{figure}

In sample 2 the absorption coefficient $\Gamma$ is $\simeq$15 dB/cm at room temperature and also increases with temperature
decrease. Figure~\ref{fig2} shows the dependencies of absorption  and variation of SAW velocity in the absence of magnetic field in sample 2.

\begin{figure*}
\centerline{
\includegraphics[width=16cm,clip=]{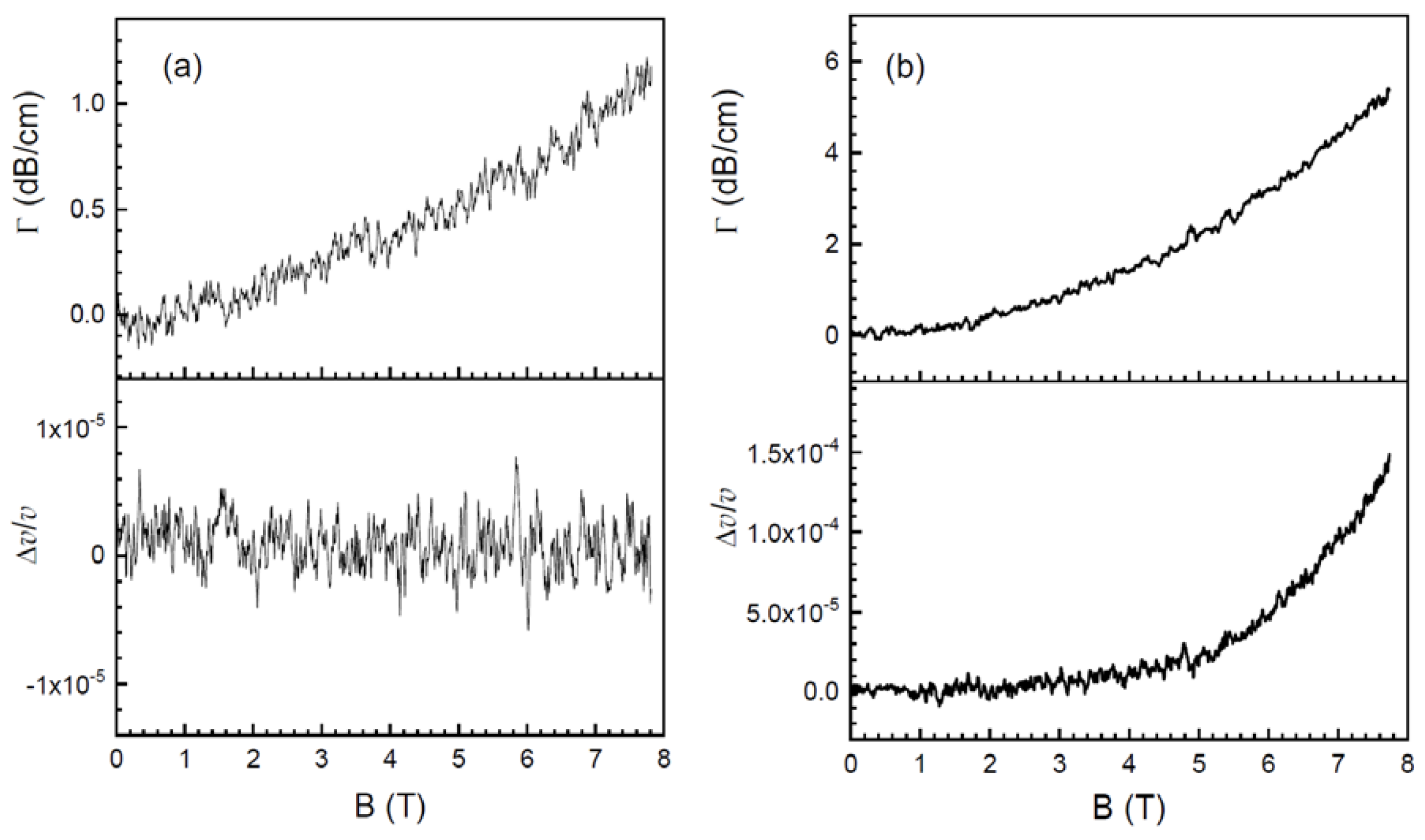}}
\caption{(a) is the dependence of $\Gamma$ and $\Delta v/v$ on magnetic field $B$; $f$ = 140 MHz, $T$ = 1.7 K; sample 1. (b) is the dependence of $\Gamma$ and
$\Delta v/v$ on magnetic field $B$; $f$ = 85 MHz, $T$ = 1.7 K; sample 2.
} \label{fig3}
\end{figure*}

Figure~\ref{fig3} shows the dependencies of $\Delta\Gamma = \Gamma(4.2 K)
- \Gamma(300 K)$ and $\Delta v/v$ on B for sample 1.

As shown in the figure, the value of SAW electronic absorption for sample 1 in magnetic field of 8 T is very small and does not exceed value of 1.2 dB/cm, while SAW velocity almost does not depend on $B$. Values of $\Gamma$ and $\Delta v/v$ for sample 2 are significantly bigger than for sample 1.

For electrical characteristics determination we use the procedure of experimental results processing, described above.

\subsection*{Sample 1}

\begin{figure}[h]
\centerline{
\includegraphics[width=9cm,clip=]{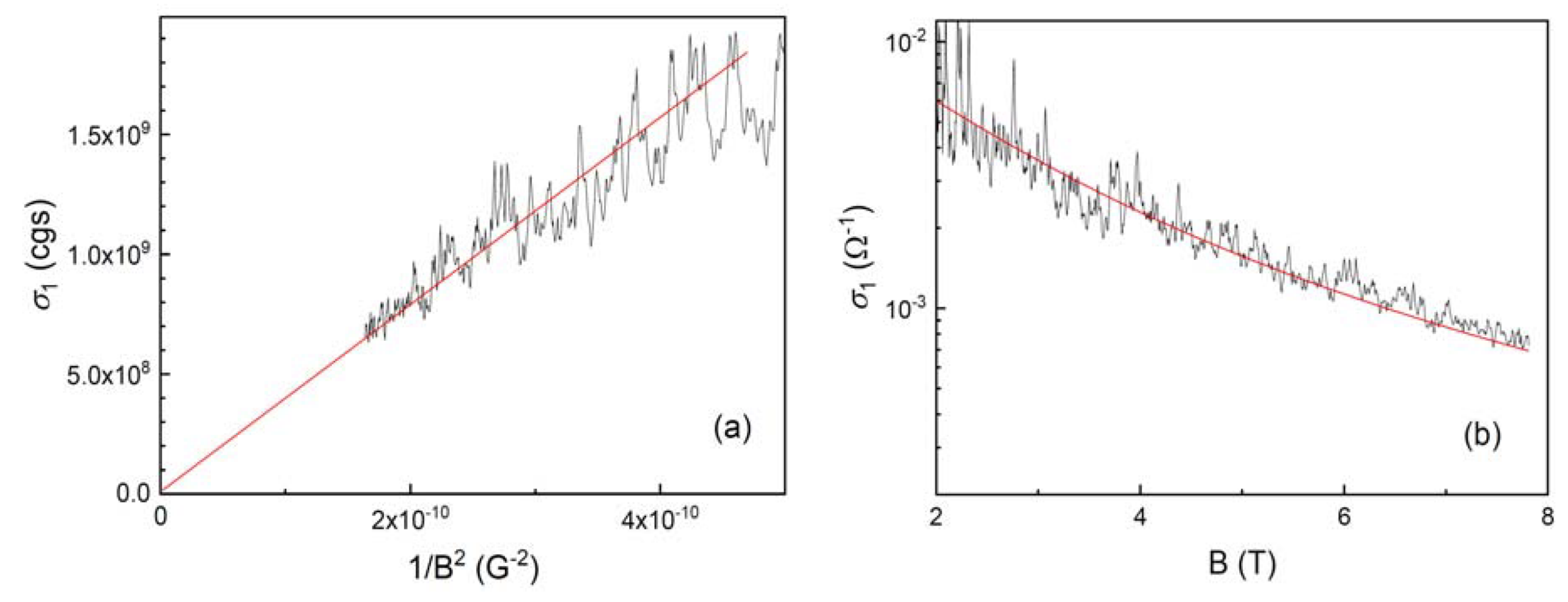}}
\caption{(a) is the dependence of $\sigma_1$~(CGS) on $1/B^2$~(G$^{-2}$);
(b) is the dependence of $\sigma_1$ ($\Omega^{-1}$) on $B (T)$ for sample 1, $f$=140~MHz, $T$=1.7~K}
\label{fig4}
\end{figure}

Figure~\ref{fig4}, (a) shows the dependence of experimentally defined $\sigma_1$ on $1/B^2$. In the area of magnetic field of 8-5 T this dependence is linear with slope $d \sigma /d(1/B^2)$=4$\times 10^{18}$
($\sigma$ — in CGS units, while $B$ — in G).

For sample 1 the following characteristics were observed using formulas~\ref{muc},~\ref{s0},~\ref{n} $\sigma (B=0)$=0.013~Ohm$^{-1}$,
$\mu = 5.4 \times  10^3$~cm$^2$/Vs, $n = 1.5 \times 10^{13}$~cm$^{-2}$.

Figure~\ref{fig4}, b shows the dependencies of experimentally defined conductivity on magnetic field (black) and calculated as per formula ~\ref{sB} (red) with values presented above. It is
shown that the red curve describes the experiment well.

\subsection*{Sample 2}

\begin{figure}[h]
\centerline{
\includegraphics[width=9cm,clip=]{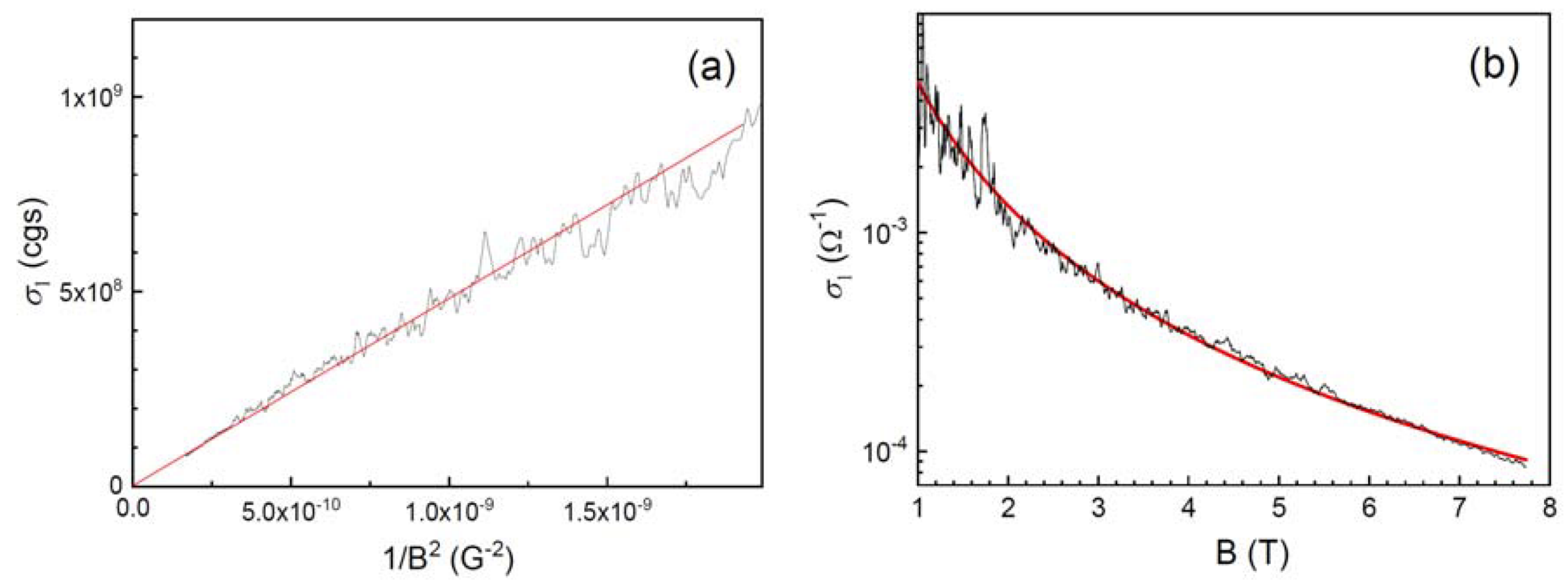}}
\caption{(a) is the dependence of $\sigma_1$~(CGS) on $1/B^2$~(G$^{-2}$);
(b) is the dependence of $\sigma_1$ ($\Omega^{-1}$) on $B (T)$ for sample 2, $f$=85~MHz, $T$=1.7~K}
\label{fig5}
\end{figure}

Figure~\ref{fig5}, a shows the experimental dependence of $\sigma_1$ on $1/B^2$. In the area of magnetic field of 8-2 T the dependence
is linear with slope in $d \sigma /d(1/B^2)$=4.9$\times 10^{17}$ ($\sigma$ — in CGS units, while $B$— in G). For sample 2 the following characteristics (1.7 K, 85 MHz) were observed using the formulas~\ref{muc},~\ref{s0},~\ref{n} :
\begin{equation}
\begin{gathered} \nonumber
\sigma (B=0)=0.043~\text{Ohm}^{-1},\ \\
\mu=2.8\times  10^4~\text{cm}^2/\text{Vs},\ \\
n=1 \times 10^{13}~\text{cm}^{-2}.
\label{dat}
\end{gathered}
\end{equation}

Figure~\ref{fig5}, b shows the experimental dependence of conductivity on magnetic field  (black)  and  calculated  (red) as per formula (2) with values defined using the method, presented above. It is shown that the red curve with these
parameters describes the experiment well.

\begin{figure}[h]
\centerline{
\includegraphics[width=9cm,clip=]{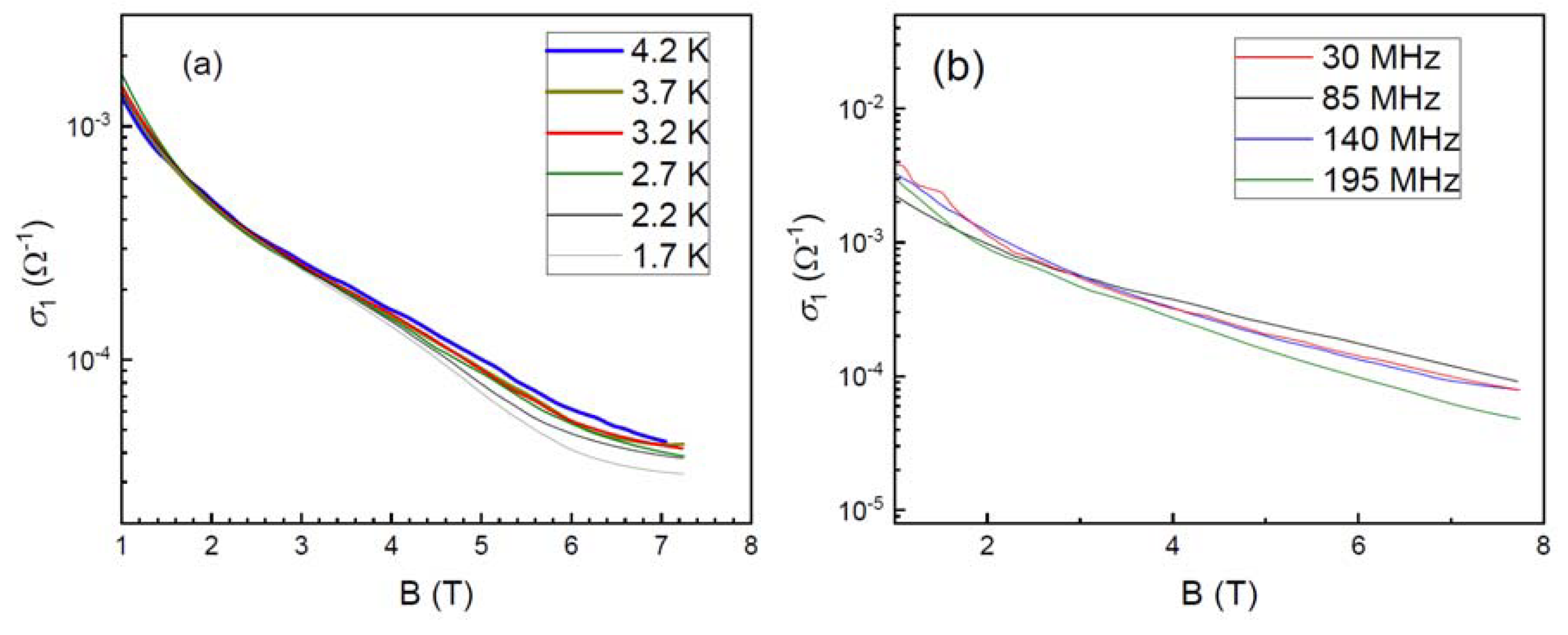}}
\caption{(a) is the dependence of $\sigma_1$ on magnetic field $B$ at various temperatures, $f$=140 MHz;  (b) is the dependence of $\sigma_1$ on $B$  for  various frequencies of SAW at $T$ = 1.7 K. }
\label{fig6}
\end{figure}

Figure~\ref{fig6} illustrates experimental dependencies of conductivity $\sigma_1$ on magnetic field B at various temperatures, measured at frequency $f$=140 MHz (a), and dependence
of $\sigma_1$ on $B$ for various SAW frequencies at $T$=1.7 K (b).

The figures show that conductivity weakly depends
on  $T$  and  $f$ ,  while  the  average   values   for   temperature  dependence  in  magnetic  field  of  7 T  are  equal  to  $\sigma_1=(4.0 \pm 0.3) \times 10^{-5}$~Ohm$^{-1}$, while for frequency dependence $\sigma_1=(7.6 \pm 1.0) \times 10^{-5}$~Ohm$^{-1}$ within measurement
error.	However,   it  can  be  seen  that  the  average   values
differ  almost   by factor 2. The  reason  is  that  measurements
of temperature dependence were made in a single day (within single cooling cycle), while the frequency dependencies — in different days (within different cooling cycles).

\begin{figure}[h]
\centerline{
\includegraphics[width=9cm,clip=]{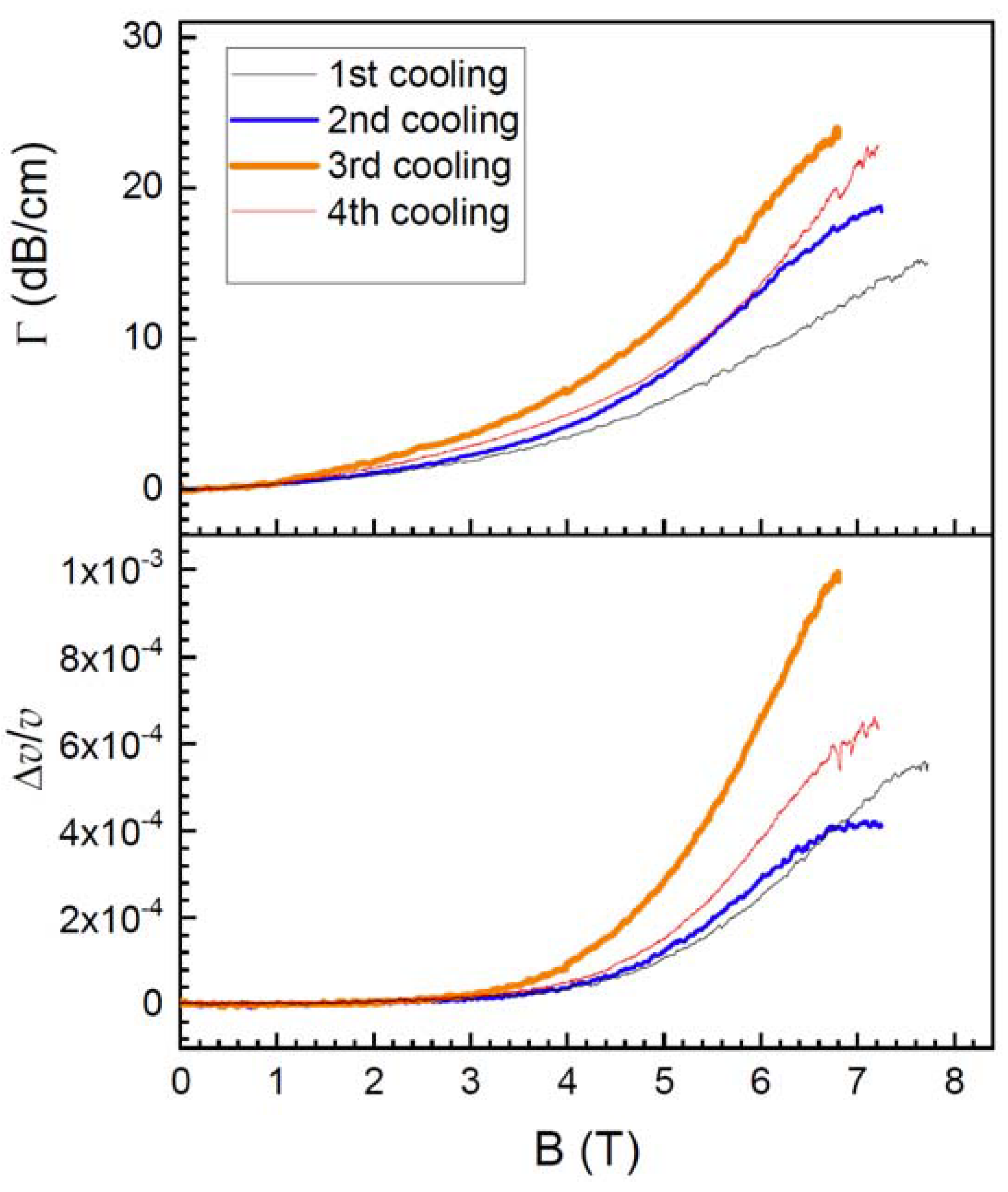}}
\caption{Dependence  of  $\Gamma$ and  $\Delta v/v$ on  magnetic  field  $B$; $f$=140 MHz, $T$=1.7 K at various cooling cycles.}
\label{fig7}
\end{figure}

We observed that with each new cooling cycle for graphene sample, deposited on lithium niobate, the dependencies of values of absorption and variation of sound velocity  on  magnetic  field  were  not  reproduced.    This is
shown in Figure~\ref{fig7} for $f$=140 MHz and $T$=1.7 K.

Calculation using formulas (3)-(5) showed  that conductivity $\sigma (0$), defined from these curves, is within a range from $4 \times 10^{-3}$ to 1.3$\times 10^{-2}$~Ohm$^{-1}$, mobility $\mu$ is within a range from 9.4$\times 10^3$ to 1.8$\times 10^4$~cm$^2$/Vs, while concentration $n$
is from 2$\times 10^{12}$ to 4.5$\times 10^{12}$~cm$^{-2}$.

\subsection{Atomic force microscopy (AFM)}

\begin{figure}[h]
\centerline{
\includegraphics[width=9cm,clip=]{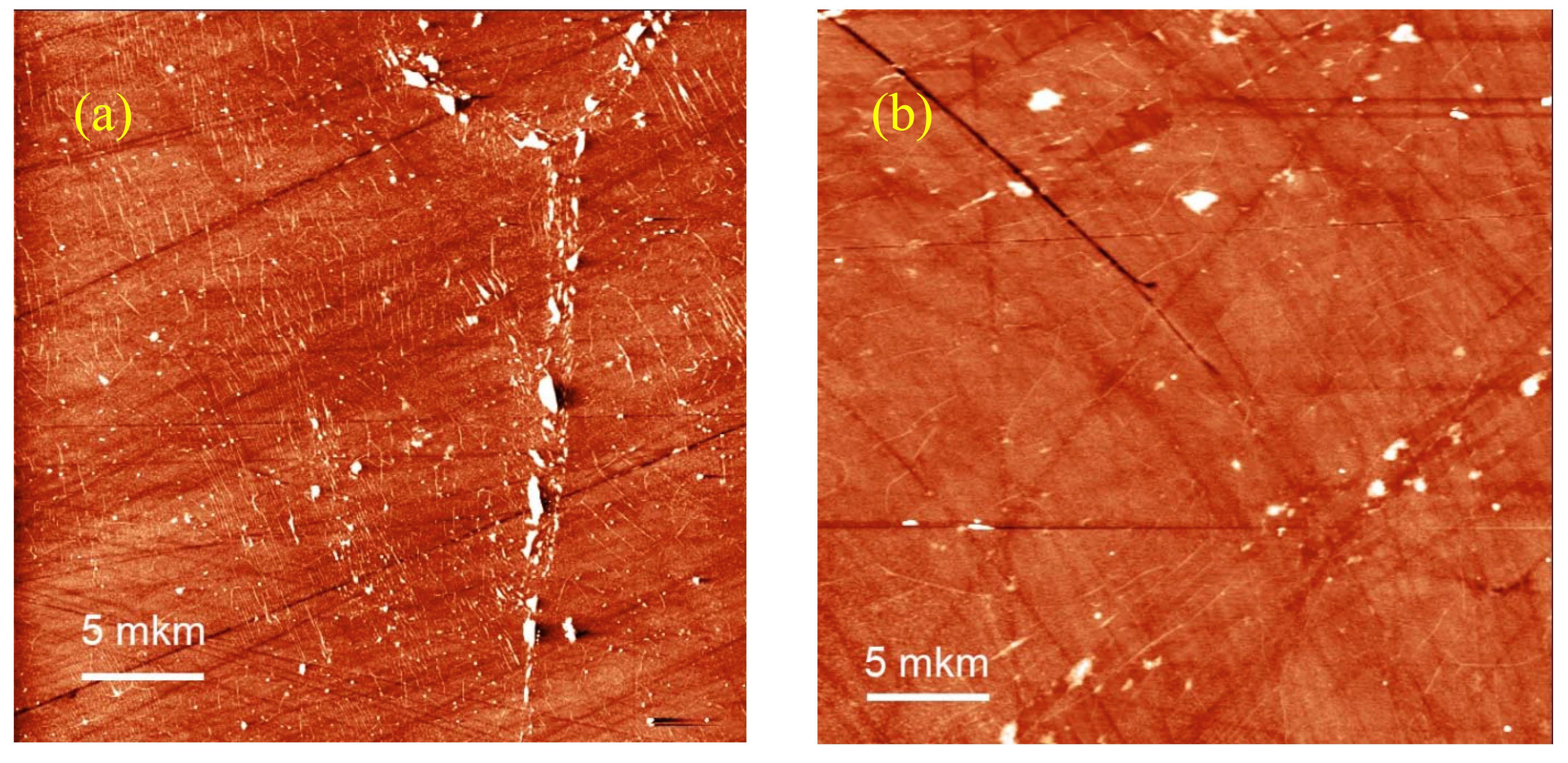}}
\caption{Topography of areas of 30$\times$30~$\mu$m$^2$ for sample 1 (a) and sample 2 (b).}
\label{fig8}
\end{figure}

Samples were studied using methods of atomic force and kelvin-probe microscopy. The main attention was drawn to
areas of relatively large 30$\times$30~$\mu$m$^2$ size. Typical images for samples 1 and 2 are presented in Figure~\ref{fig8}.

\begin{figure}[h]
\centerline{
\includegraphics[width=9cm,clip=]{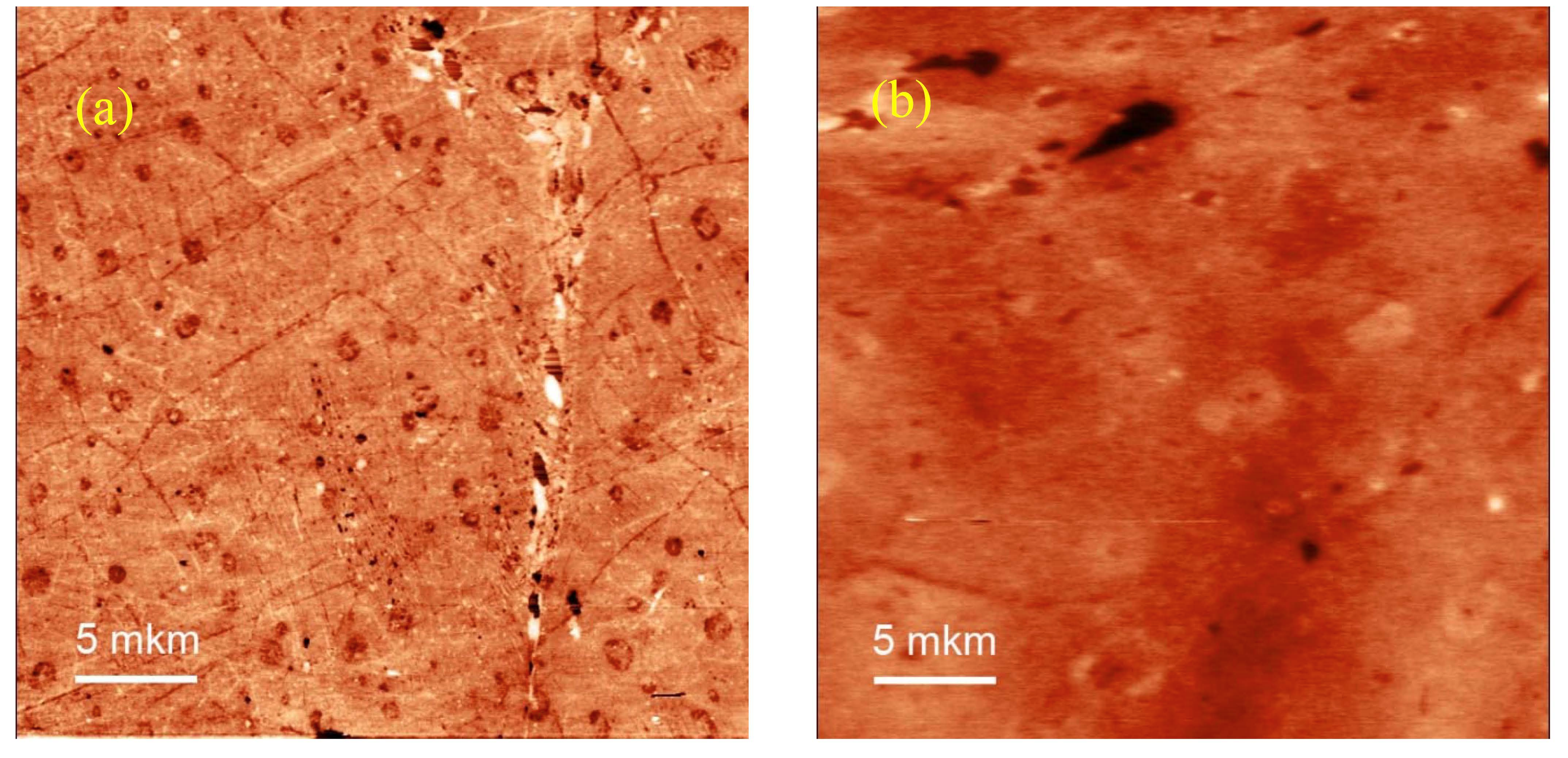}}
\caption{Distribution of surface potential of areas of 30$\times$30~$\mu$m$^2$ for sample 1 (a) and sample 2 (b).  The darker color shades correspond to lower values of surface potential.}
\label{fig9}
\end{figure}

It is well seen that sample 1 is covered with a  large amount of linear defects, which probably represent graphene wrinkles, appeared after sample cooling and heating cycles. Such peculiarities are also seen on sample 2, but in significantly less amount. The specified difference of the samples is more noticeable in potential distribution images
(Figure~\ref{fig9}).

Two types of defects are shown on sample 1. Extended
peculiarities, combined into the common network, probably correspond to grain boundaries of polycrystal graphene. Also, the round peculiarities are observed, that we associate with islands of the second graphene layer. It should be noted that the majority of linear objects, observed on topography image, is not reflected in surface potential distribution.

Potential distribution on sample 2 is much more homogeneous. Several areas with reduced potential, that we also
associate with the islands of  the  second  graphene  layer,  are observed. Drop-like defect, observed in the upper part, corresponds to rupture in graphene layer.

\section{Discussion of results}

Let’s start the results discussion with  dependence  of SAW absorption coefficient $\Gamma$ on  temperature  at  $B$=0.  In study \cite{8} it  was  specified  that  for  monolayer  undoped graphene, deposited on SiO$_2$/Si, resistance at
room temperature in the absence of magnetic field  was about 300-1000 Ohm per  square.  With  such  resistance the  SAW  absorption  with  charge   carriers  is  very   low,   $\Gamma <$~1~dB/cm, therefore the high SAW absorption, observed in the experiment, and its weak dependence on temperature probably  indicate  the  absorption  (scattering) of SAW  with
structure defects in polycrystal film of graphene. Absorption
in sample 1 was much higher than in sample 2, that is presumably caused by the fact that sample 1  has  much  more defects than sample 2. This is confirmed with the results of samples study using the methods of atomic force and kelvin-probe microscopy at room temperature.

As for Figs.~\ref{fig8} and ~\ref{fig9}, the individual graphene grains as part of polycrystalline film are well-visualized on sample 1. Also, the wrinkles are observed on surface due to sequential processes of sample cooling and heating. Despite the same polycrystalline nature of sample 2, no grain boundaries are observed on it.

We suppose that increase of absorption (scattering) of SAW    during   sample   cooling   is   related   to   increase of
amount of  defects,  for instance,  ruptures  in film,  that   are
"cured" during sample heating after measurements. Increase
of defects concentration during samples cooling happens randomly, as per Fig.~\ref{fig7}.

Dependence of $\sigma_1$ on magnetic field, temperature and frequency indicates the metal nature of conductivity at low temperature, while in sample 1 with higher amount of defects the carriers mobility is less than in sample 2. Relation $\sigma_1 > \sigma_2$ also indicates the metal nature of conductivity.

It turned out that values of charge carriers mobility in graphene (5400 and 28000 cm$^2$/Vs), obtained using non-contact acoustic method, are much higher than mobility values (1100-1500 cm$^2$/Vs), obtained by the measurement of the  volt-ampere  characteristics (VAC) of  graphene
field transistor (GFT), made of similar samples~\cite{11,12}. It seems,  it  can  be  explained  with  several  factors.	Firstly,
the measurements of volt-ampere characteristics in GFT were  made  at  room  temperature.  Secondly,  preparation for these measurements assumes multiple lithography and polymer application, as well as other process manipulations with samples, that result in  their  degradation.  Thirdly, when measuring the VAC in GFT, graphene contacts with metal, that can significantly change the electronic structure of graphene and make contributions to its charge carriers mobility.

\section{Conclusion}

Non-contact   acoustic   methods   use    for    studying   the low-temperature electrical properties  of  monolayer CVD graphene, deposited on lithium niobate surface, allowed to define its new characteristics. These characteristics
include variation (increase) of amount of defects at samples cooling from room temperature to 4.2 K and random nature
of  these  defects  formation  at  various  cooling  cycles. The
latter property expressed in limited reproducibility of the measured absorption and variation of SAW velocity. Along with determination of the material electrical characteristics, the mechanism of low-temperature conductivity was defined.

\paragraph*{Acknowledgments}
The study was supported by the Russian Foundation for Basic Research grant  19-02-00124.  Graphene  samples were prepared in the Institute of General Physics of the Russian Academy of Sciences within the framework of studies as per grant of the Russian Science Foundation
 21-72-10164.

\vfill\eject

\end{document}